\documentclass[namedreferences]{kluwer}

\usepackage{graphicx}

\newcommand{\cago}{$^{12}{\rm C}(\alpha,\gamma)^{16}{\rm O}$\ }

\begin{document}
\begin{article}

\begin{opening}         

\title{The \cago Nuclear Reaction Rate from \\ 
       Asteroseismology of the DBV White Dwarf CBS~114}

\author{T. S. \surname{Metcalfe}}
\institute{Theoretical Astrophysics Center, Aarhus University, Denmark}

\author{G. \surname{Handler}}
\institute{South African Astronomical Observatory, Observatory 7935, South
Africa}

\runningauthor{Metcalfe \& Handler}
\runningtitle{\cago from Asteroseismology of CBS~114}

\begin{abstract}
We have identified seven independent pulsation modes in the
helium-atmosphere variable (DBV) white dwarf star CBS 114, based on 65
hours of time-resolved CCD photometry from the 0.75-m telescope at SAAO.  
We interpret these pulsations as non-radial g-modes with the same
spherical degree $\ell$=1, as suggested by the mean period spacing of
$37.1\pm0.7$ seconds. We use a genetic-algorithm-based fitting method to
find the globally optimal model parameters, including the stellar mass
($M_*=0.73\ M_{\odot}$), the effective temperature ($T_{\rm
eff}=21,000$~K), the mass of the atmospheric helium layer ($\log[M_{\rm
He}/M_*]=-6.66$), and the central oxygen mass fraction ($X_{\rm O}=0.61$).
The latter value implies a rate for the \cago reaction near
$S_{300}=180$~keV~b, consistent with laboratory measurements.
\end{abstract}


\end{opening}

\section{Astrophysical Context}

Almost every star in our galaxy will eventually become a white dwarf.
Since they are relatively simple compared to main-sequence stars, white
dwarfs provide one of the best opportunities for learning about stellar
structure and evolution. The helium-atmosphere variable (DBV) white dwarfs
are among the simplest of all, with a surface helium layer surrounding a
degenerate C/O core. The internal composition and structure of white
dwarfs formed through single-star evolution is largely determined by the
relative rates of two nuclear reactions that compete for the available
helium nuclei during core helium burning in the red giant phase of
evolution: the $3\alpha$ and \cago reactions. The $3\alpha$ rate is
well-constrained from laboratory measurements, but the \cago rate is still
very uncertain. Recent advances in the analysis of asteroseismological
data on DBV stars \cite{mwc01} now make it possible to obtain precise
measurements of the central C/O ratio, providing a more direct way to
determine the \cago rate at stellar energies. The first application of
this new method to the star GD~358 implied a reaction rate that is
significantly higher than most extrapolations from laboratory data
\cite{msw02}. Fortunately, each pulsating white dwarf can provide an
independent measurement of the reaction rate, so the analysis of
asteroseismological data for additional DBV stars would be useful.

\section{Observations}

The pulsations of the DBV star CBS~114 were discovered by
\citeauthor{wc88}~(\citeyear{wc88}; \citeyear{wc89}), who reported
multiperiodic oscillations with peak-to-peak light variations up to 0.3
mag with a time scale of about 650 seconds. No further observations of
this star have been published, so we decided to obtain a larger data set
to study the pulsations of CBS~114 in more detail. We acquired 65 hours of
differential CCD photometry with the 0.75-m telescope at the Sutherland
station of the South African Astronomical Observatory during three weeks
distributed over a period of two months in the spring of 2001.

\begin{figure}
\centerline{\includegraphics[width=12truecm]{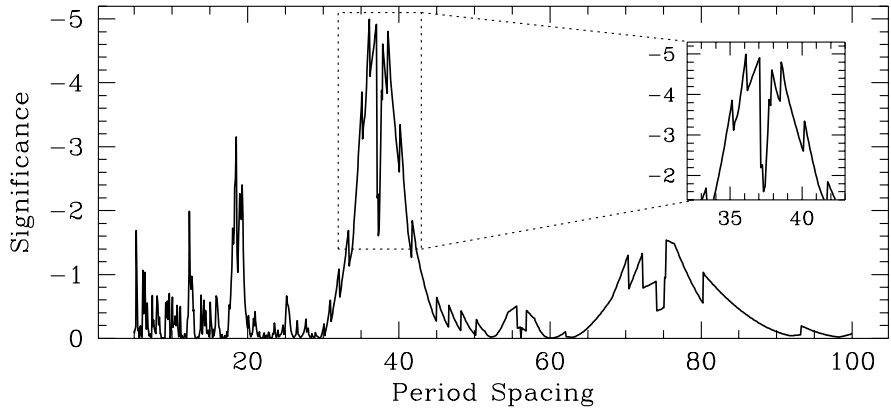}}
\caption[]{The significance of various values for the mean period spacing 
as determined by the Kolmogorov-Smirnov test for the pulsation periods 
observed in CBS~114. The diagram is dominated by peaks near 40 s and 
their (sub)harmonics, implying a most probable identification of $\ell$=1 
for the modes. Conspicuously absent are any peaks near 22 s, the expected 
spacing for $\ell$=2 modes.} 
\label{fig1} 
\end{figure}

The amplitude spectrum of our combined data set shows six dominant
structures between 1.4 and 2.6 kHz. We found these same frequencies in the
discovery data of \citeauthor{wc88}, which contained a seventh frequency
that was also present in our data, but near the detection threshold. Thus
our final multifrequency solution contained seven independent modes
\cite{hmw02}. We determined the mean period spacing of these modes by
calculating the power spectrum of the period values with unit amplitude
\cite{han97}, leading to a value of $37.1\pm0.7$~s. This agrees well with
the results of a Kolmogorov-Smirnov test, shown in Figure~\ref{fig1}.

\section{Model Fitting}

In the asymptotic limit for a typical (0.6 $M_{\odot}$) white dwarf model,
the mean period spacing between pulsation modes of consecutive radial
overtone is near 40 seconds for $\ell$=1 and 22 seconds for $\ell$=2
\cite{bww93}. No modes with spherical degree higher than $\ell$=2 have
ever been detected in a white dwarf, but they lead to even lower values
for the mean period spacing. Thus, we interpret the modes in CBS~114 as
arising from pulsations with $\ell$=1 and m=0. Since we did not observe
any multiplet structure, the latter assumption has no justification.
However, \inlinecite{met02a} has shown that the consequences of this
assumption are not serious for the purposes of model-fitting when the
rotation period is $\gtrsim$1 day, as has been the case for other
pulsating white dwarfs.

Using the optimization method developed by \citeauthor{mnw00}
(\citeyear{mnw00}; \citeyear{mwc01}), we performed a global search for the
optimal model parameters to fit the seven independent pulsation periods of
CBS~114. The method uses a parallel genetic algorithm to minimize the
root-mean-square (rms) differences between the observed and calculated
periods for models with effective temperatures ($T_{\rm eff}$) between
20,000 and 30,000 K, total stellar masses ($M_*$) between 0.45 and 0.95
$M_{\odot}$, a helium layer mass with $\log[M_{\rm He}/M_*]$ between
$-$2.0 and $-$7.3, and a simple parameterization of the C/O profile that
fixes the oxygen mass fraction to its central value ($X_{\rm O}$) out to
some fractional mass ($q$) where it then decreases linearly in mass to
zero oxygen at 0.95 $m/M_*$. The optimal values for the 5 model parameters
were:  $T_{\rm eff}=21,000$~K, $M_*=0.73\ M_{\odot}$, $\log[M_{\rm
He}/M_*]=-6.66$, $X_{\rm O}=0.61$, and $q=0.51\ m/M_*$. The rms period
residuals of this fit were $\sigma_{\rm P}=0.43$ s, which is significantly
better than any fit where only the first three parameters are adjusted and
the core composition is either pure carbon or oxygen. We quantified the
uncertainty in the derived central oxygen mass fraction ($\Delta X_{\rm
O}=\pm 0.01$) by calculating a grid of models with various combinations of
$X_{\rm O}$ and $q$, fixing the other three parameters at their optimal
values.

\section{Results \& Discussion}

Our optimal model for CBS~114 has a higher mass and a lower central oxygen
mass fraction than the optimal model for GD~358 \cite{mwc01}.  Turning
these values into a measurement of the \cago rate requires the calculation
of evolutionary internal chemical profiles like those of
\inlinecite{sal97}. A model of the internal chemical profile with the same
mass as our fit to CBS~114 requires a rate for the \cago reaction of
$S_{300}=177\pm3$~keV~b (internal uncertainty) to match the derived
central oxygen abundance within the 1$\sigma$ limits (M.~Salaris, private
communication). This value is consistent with (but much more precise than)
the rate derived from recent high-energy laboratory measurements
($S_{300}=165\pm50$~keV~b; \opencite{kun02}). By contrast, the rate
previously derived from a similar treatment of the white dwarf GD~358 was
significantly higher ($S_{300}=370\pm40$~keV~b). However,
\inlinecite{met02b} identified a systematic difference in the analysis of
GD~358 and CBS~114. After correcting this difference in treatment, the two
sets of data yielded reaction rates that are both consistent with
laboratory measurements, and marginally consistent with each other.

\inlinecite{bf02} have proposed an alternative model to explain the
pulsation spectrum of GD~358, involving two composition transition zones
in the surface helium layer, as suggested by time-dependent diffusion
calculations similar to those done by \inlinecite{dk95}. \citeauthor{bf02}
note that their fit, which has a pure carbon core, is actually worse with
a core of either pure oxygen or a uniform C/O mixture. An important test
of their alternative model will be whether it can match the pulsation
spectrum of CBS~114 with the same set of assumptions used in the analysis
of GD~358, as \inlinecite{met02b} has done with the model used here.

\end{article}
\end{document}